# Contrast Phase Classification with a Generative Adversarial Network

Yucheng Tang[*a], Ho Hin Lee[a], Yuchen Xu[a], Olivia Tang[a], Yunqiang Chen[b], Dashan Gao[b], Shizhong Han[b], Riqiang Gao[a], Camilo Bermudez[e], Michael R. Savona[c], Richard G. Abramson[d], Yuankai Huo[a], Bennett A. Landman[a,d,e]

[a]Department of Eletrical Engineering and Computer Science, Vanderbilt University, Nashville, TN, USA 37212;
[b]12 Sigma Technology, San Diego, CA, USA 92130;
[c]Hematology and Oncology, Vanderbilt University Medical Center, Nashville, TN, USA 37235;
[d]Radiology, Vanderbilt University Medical Center, Nashville, TN, USA 37235;
[e]Department of Biomedical Engineering, Vanderbilt University, Nashville, TN, USA 37212

(*Corresponding author: yucheng.tang@vanderbilt.edu)

## ABSTRACT

Dynamic contrast enhanced computed tomography (CT) is an imaging technique that provides critical information on the relationship of vascular structure and dynamics in the context of underlying anatomy. A key challenge for image processing with contrast enhanced CT is that phase discrepancies are latent in different tissues due to contrast protocols, vascular dynamics, and metabolism variance. Previous studies with deep learning frameworks have been proposed for classifying contrast enhancement with networks inspired by computer vision. Here, we revisit the challenge in the context of whole abdomen contrast enhanced CTs. To capture and compensate for the complex contrast changes, we propose a novel discriminator in the form of a multi-domain disentangled representation learning network. The goal of this network is to learn an intermediate representation that separates contrast enhancement from anatomy and enables classification of images with varying contrast time. Briefly, our unpaired contrast disentangling GAN(CD-GAN) Discriminator follows the ResNet architecture to classify a CT scan from different enhancement phases. To evaluate the approach, we trained the enhancement phase classifier on 21060 slices from two clinical cohorts of 230 subjects. The scans were manually labeled with three independent enhancement phases (non-contrast, portal venous and delayed). Testing was performed on 9100 slices from 30 independent subjects who had been imaged with CT scans from all contrast phases. Performance was quantified in terms of the multi-class normalized confusion matrix. The proposed network significantly improved correspondence over baseline UNet, ResNet50 and StarGAN's performance of accuracy scores 0.54. 0.55, 0.62 and 0.91, respectively (p-value<0.0001 paired t-test for ResNet versus CD-GAN). The proposed discriminator from the disentangled network presents a promising technique that may allow deeper modeling of dynamic imaging against patient specific anatomies.

## 1. INTRODUCTION

Computed Tomography (CT) is widely used in clinical applications and offers standardized Hounsfield Units (HU) for quatitative tissue comparison. In the past decades, contrast enhanced CT has a played important role of providing imaging contrast, characterizing organ physiology, and detecting lesions[1]. When acquiring contrast enhanced CT, iodine contrast material is injected into a peripheral vein passes through central vein, atrium in heart, pulmonary vein, lung, arterial system, liver, spleen, pancreas, kidney and urinary systems. The CT scanner is timed to acquire one or more scans during the enhancement cycle [2]. However, the phase information is typically recorded by manually. Therefore, mislabeling can happen among large-scale cohorts, which is resource intensive to detect and correct. Moreover, the phase label correction requires the human to differentiate contrast information from enhancement phases, which is hard due to variations in tissue, contrast material, injection protocols, vascular dynamics and metabolism variance. Therefore, the ability to automatically classify clinical scans has potential value for contrast evaluation and assist anatomical image processing.

Herein, we target detection of three typical contrast enhancement phases (1) non-contrast, i.e, without any contrast material effect. (2) portal venous phase, i.e., acquired 70 to 80 seconds after bolus injection. In this phase, the heart is slightly

bright,while the aorta, liver and spleen are even brighter. (3) Delayed, i.e., acquired 10 minutes after bolus injection. In delayed scans, contrast materials have diffused through the urinary system and outer area of kidneys, while other parts of the image are relatively dark. In summary, the purpose of above enhancements is to characterize pathology and physiology of organs across different enhance stages. The examples are shown Figure 1, which illustrates how contrast enhancements are distributed, and why some slices are hard to classify.

In the past decades, image analyses on contrast enhanced CT, especially deep learning based methods, have attracted increasing attention [3]. One of the most fundamental challenge is to perform contrast phase classification, with the aim to accurately classify an unlabeled contrast CT with a phase label as an image classification task [4, 5]. Traditionally, the image classification model is supervised by statistical learning-based loss functions (e.g., margin loss and cross entropy loss). Adversarial learning frameworks [6, 7] have been introduced to the classification tasks by training a data-driven discriminator instead of traditional loss functions. Meanwhile, the generative adversarial network (GAN) is able to perform domain adaption when introducing a generator. More recently, multi-domain discriminator [8] is presented to solve classification problem while improving performance of synthesis.

In this paper, we propose a multi-domain contrast disentangling GAN (CD-GAN) to perform multi-phase contrast enhanced CT phase classification, by learning disentangled representations across contrast phases features. The method (Figure 2. (d)) takes 2-D slices to learn an intermediate representation, then reconstruct a synthetic contrast enhanced image. The goal of using generated image is to employ adversarial loss and synthetic classification as data augmentation for improving discriminator's performance.

Our contribution in this study are as follows:
- We propose CD-GAN, an adversarial learning network to perform contrast phase classification by learning from both real and synthetic images.

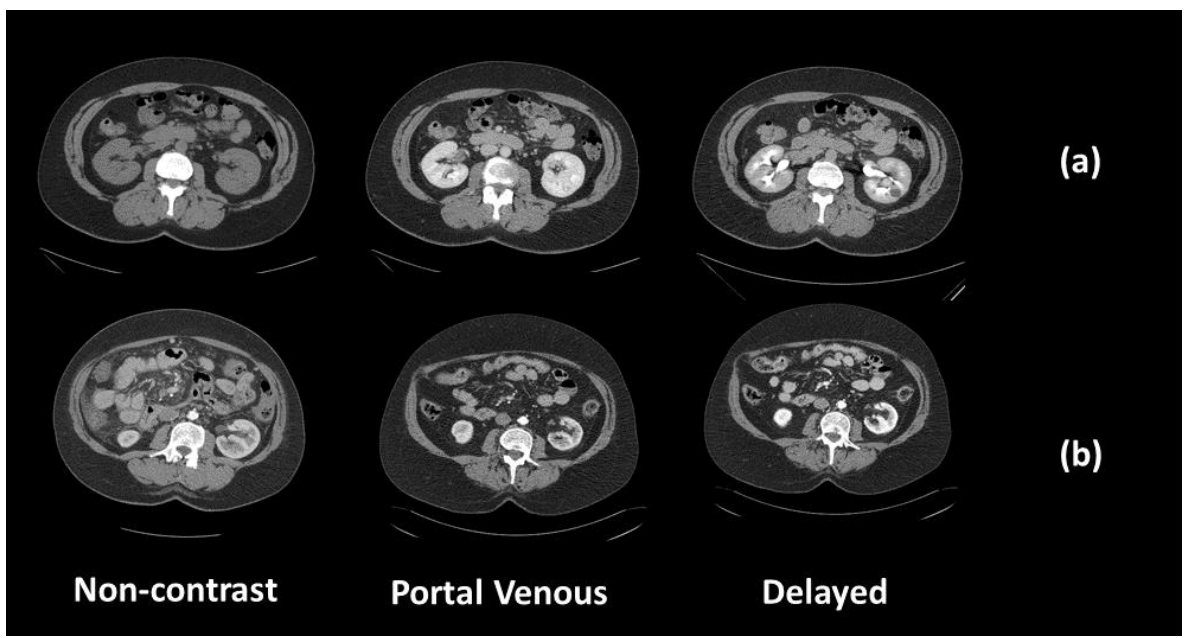

Figure 1.The challenge of classification among contrast enhancements with row a) slices with standard enhancement phase; b) slice with large variance in different tissues due to contrast protocols, vascular dynamics, and metabolism.

- A two-step generator is introduced, which learns the disentangled representations and domain specific features among multiple contrast CT.
- We perform quantitative and qualitative evaluations on of the prevalent baselines (UNet, ResNet50 and StarGAN), and compared the performance with the proposed CD-GAN.

230 volumes with 21060 2-D CT slices are used in this study. From the results, the proposed CD-GAN achieve superior overall accuracy score (0.91) over U-Net, ResNet50 and StarGAN (0.54. 0.55, and 0.62), respectively.

## 2. METHODS

In this paper, we propose a discriminator from contrast disentangling generative adversarial network (CD-GAN), presented in the rightmost panel of Figure 2. We first describe our disentangling representation learning framework for proposed discriminator. Then, we discuss how adversarial loss and real/synthetic images classification help boost classification performance.

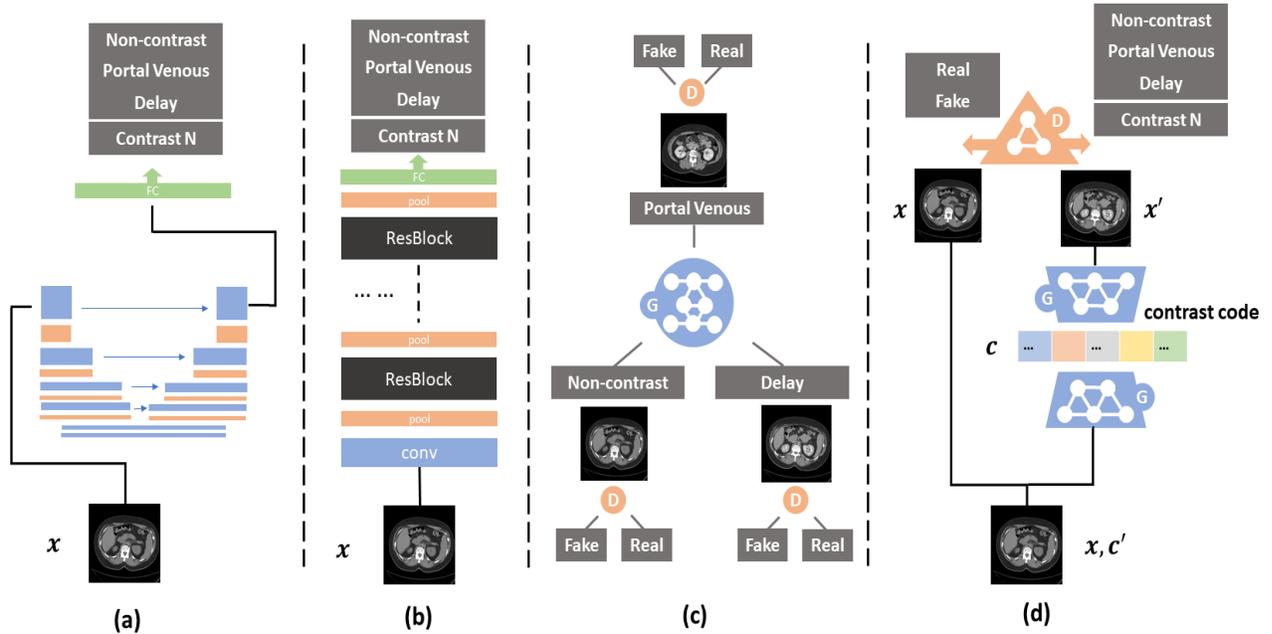

Figure 2. Baseline model UNet (a), ResNet (b), StarGAN (c) and our proposed method of discriminator in Contrast Disentangling GAN (CD-GAN) (d). The proposed classifier uses both adversarial loss and contrast classification loss on real and synthetic image.

### 2.1 Prior work

**Contrast Enhancements:** Contrast enhanced CT studies are widely used in diagnostic analysis and involve litter physiological stress [1]. There are several basic enhancement protocols: non-contrast, early arterial, late arterial, hepatic, portal venous and delayed phases which are acquired by different timing orders [2]. In our work, we use three most typical phases: non-contrast, portal venous and delayed phases as the problem framework.

**Phase detection**: Recently, qualitative and quantitative evaluation studies are presented in phased-contrast CTs [9]. Detecting phase and optimal timing for injection shows potential value of identifying lesion and abnormal tissues [10, 11].

**Deep Classification model:** Deep convolutional neural networks have been presented to a series of breakthroughs for image classification [12]. With the ImageNet dataset, ResNet revealed the levels of features can be stacked to an end-to-end classifier figure 2 (left). The residual learning framework provides reasonable preconditioning on classification tasks.

**Generative Adversarial Network:** GAN have been achieved remarkable achievement on computer vision problems like recognition [13], image translation [14], style transfer [15] and super-resolution [16]. A typical GAN model has two adversarial process. With a minimax two controversial player games, the generator and discriminator leverage each other.

**Multi-domain Discriminators**：The task for discriminators is to distinguish a real image from a synthetic one [6]. However, discriminator can be assigned with more complicated tasks. DR-GAN built a model that G try to fool D to classify input to the identity domain and face poses. StarGAN presented an approach that learn classification feature from both real and synthetic images [17]. While GauGAN [15] proposed a framework with multi-modal synthesis and style-guided transfer [15]. We explored StarGAN as the baseline of representative GAN approach.

## 2.2 Discriminator in Contrast Disentangling Generative Adversarial Network (CD-GAN)

**Adversarial Loss in Discriminator:** Prior GANs proposed adversarial loss to synthetic samples from empirical image and convolutional neural networks. In our tasks, the discriminator is not mainly to distinguish the real and synthetic images, Therefore, the idea of using adversarial loss is to boost classification performance in D by performing adversarial learning, rather than rely on canonical loss functions and regularities. Our rationale is to use adversarial learning framework to supervise the training process and make classification more robust by alleviating spatial inconsistency and leveraging overall performance. For distinguishing real and synthetic images and perform the two-play minmax game, we adopt the adversarial loss:

$$\mathcal{L}_{adv} = E_x[logD(x)] + E_{x,c}[\log(1 - D(G(x,c))] \quad (1)$$

Given a real input contrast enhancement slice $x$. $G$ generates an image $G(x,c)$ conditioned on input contrast enhancement image and target contrast phase code $c$. $D$ tries to discriminate real, synthetic image and maximize the objective: $\min_G \max_D \mathcal{L}_{adv}(G,D)$.

**Contrast Classification Loss in Discriminator**: For a given contrast phase slice $x$, along with its target contrast code $c'$, the main goal of discriminator is to classify real image $x$ and synthetic image $x'$ to its proper categories. To achieve this condition, we designed an auxiliary classification loss on top of $D$ when maximizing $D$ and minimizing $G$. We use contrast classification loss on both real and synthetic images to optimize $D$. For real image path, the contrast classification loss is defined as:

$$\mathcal{L}_{cls} = E_{x,c'}[-\log(1 - D_{cls}(c'|x))] \quad (2)$$

This term represents the probability distribution over its true contrast phase labels, where $D$ tries to learn to classify real scan slices to its contrast target label. Also, for synthetic image path, we define the contrast classification loss as:

$$\mathcal{L}_{cls'} = E_{x,c}[-\log(1 - D_{cls}(c|G(x,c)))] \quad (3)$$

where $c$ is the assigned contrast code of the generated contrast enhanced slice. $D$ tries to classify the synthetic image to its assigned condition $c$.

**Contrast Disentangling Generator:** The synthetic images, although not the central task of this network, are obtained from the generator. Using the synthetic images, our goal is to boost classification performance of discriminator. The generator in CD-GAN differently from prior GAN models with two novel designs. First, it reconstructs a contrast enhancement representation by an encoder-decoder architecture. As there are commonly distractive variations in patients' scans, we aim to generate a discriminative representation by encoder. To remedy this, we use variation information such as tissues volumes to explicitly disentangle representations. Herein, the encoder tries to disentangle the enhancement information to an intermediate representation. Second, the decoder learns the enhancement details and tries to add different contrast phase knowledge to the representation. In detail, $G$ consists of an encoder $G_{enc}$ and a decoder $G_{dec}$. $G_{enc}$ aims to learn an intermediate representation from an input contrast enhanced slice $x$ as $f(x) = G_{enc}(x)$. $G_{dec}$ aims to synthesize a synthetic contrast enhanced slice with the input of representation with its assigned contrast code $c$: $x' = G_{enc}(f(x), c)$.

**Network Structure:** We adopted the DR-GAN [18] for $G_{enc}$ where batch normalization and ReLU are employed after each convolutional layer. The intermediate representation is from AvgPool output in our network. The representation is concatenated with contrast code $c$. For high resolution image synthesis, we adopt 2D-UNet as $G_{dec}$. Also, $D$ is designed on top of PatchGAN [17] with two more convolutional layers and LeakyReLU. Finally, we add a fully connected layer with softmax function for three contrast phase categories.

**Full Objective:** $D$ and $G$ improves each other during iterative training. We seek a better classifier from D, to distinguish real and synthetic images while classifying contrast phase. $G$ os designed to synthesize a usable contrast preserving image. Overall, there are three benefits. First, using an independent contrast code for $G_{dec}$, $G_{enc}$ is trained to disentangle the large variance from patients. Second, the discriminator on real/ synthetic guides the contrast classification. Third, the adversarial loss help discriminator to converge.

### 2.3 Datasets

We performed de-identified data retrieval from the ImageVU project of the Vanderbilt University Medical Center (VUMC) under IRB approval. In training process, we use 230 subjects (21060 2-D slices). These subjects are independent and unpaired with three contrast phases, 72 from delayed cohort, 78 belong to portal venous phase and 80 in non-contrast phase. During testing, 30 subjects with paired scans from all three-contrast phase (90 scans in total) are used as testing data, whose data are acquired from different studies along with different contrast protocols. For pre-processing, we performed body part regression [19] to separate abdomen area from shoulder and pelvis in whole body CT scan. Then we converted the abdomen volume to axial view 2-D slices while preserving the original resolution by 0.8x0.8mm with dimension 512x512. For consistency, we employ a wide HU window from -1000 to 1000 to remove intensity outliers. Then image intensities are linearly scaled to -1 to 1.

### 2.4 Experiment design

For baseline models (Figure 2), we adopt UNet, ResNet [4, 5], and StarGAN [8] discriminators. These models are widely used in medical image analysis, classification and multi-domain attribute transfer. All experiments are implemented with same data configuration.

**UNet:** UNet is widely used in image segmentation tasks, it enables extracting feature maps from tiny structural area. UNet presented high performance pixel-wised classification with limited medical images. We used UNet as a baseline whether extracting contrast features is a difficult task. We add a fully connected layer to the end of UNet and employed cross entropy loss as classification error function.

**ResNet50:** ResNet has presented its success in most classification problems. ResNet solved degrading problem when deeper networks start converging by using Resblocks, shown in figure 2. (left). ResNet gradually restores skip-connections and learns feature maps concisely. As we are using high resolution images, we choose ResNet50 which is a 50 layers Residual network as our baseline method. We perform same batchsize 8 in CD-GAN method, and learning rate 0.0001. Cross entropy loss is employed for multi domain classification.

**StarGAN Discriminator:** StarGAN used a single generator that learns mappings among multiple domains, while discriminator employed classification loss. The generator composed of two convolutional layers with stride size of two and six residual blocks, followed by two transposed convolution layers, discriminator adopted PatchGAN. Instance normalization is used and no normalization in discriminator. We trained a StarGAN as another comparison and presented its multi-domain discriminator's performance.

**CD-GAN Discriminator:** To provide contrast transfer labels, we apply each slice three target contrast codes include the current phase. The generator tries to recontract three enhancement types for each image. We follow the optimization strategy in [20], the batch size is set to be 8. All weights are trained from scratch. Adam optimizer is used with learning rate of 0.0001 and beta1=0.5 and beta2=0.999. We implement learning rate decay of step 1000 every 100000 iterations. Training continues with 500000 iterations and complete time of about 36 hours on single NVIDIA 2080TI GPU with 11G memory. Prior GANs suggest alternating between k steps of optimizing discriminator and one step of optimizing generator, typically k is set larger than 1 [21]. This helps discriminator to maintain optimal solution while the generator converges slowly. In our experiment, k is set to 1 due to the more complicated discriminator structure, as it presents strong supervisions for both real, synthetic images and classification labels.

### 2.5 Metric

We perform multi contrast phase classification performance with standard normalized confusion matrix, and all experiments are tested on our paired withheld data with the accuracy score.

$$accuracy(p, p') = \frac{1}{N} \sum_{i=0}^{N-1} 1(p, p') \qquad (4)$$

where, $p$ is the predicted value, while $p'$ is the corresponding real value, $N$ is the total number samples. The fraction of prediction values over $N$ is defined as the classification accuracy score.

# 3. RESULTS

## 3.1 Quantitative evaluation

We compared our method with baseline models UNet, ResNet50 and StarGAN discriminators. Figure 3 presents the confusion matrix of each model. StarGAN has similar accuracy with UNet, as StarGAN used shallower model for both generator and discriminator, that does not support high resolution image with large dimensions. ResNet provides best baseline result with the deeper 50 layers of residual network. CD-GAN discriminator benefits from synthetic image in classification tasks as well as adversarial loss.

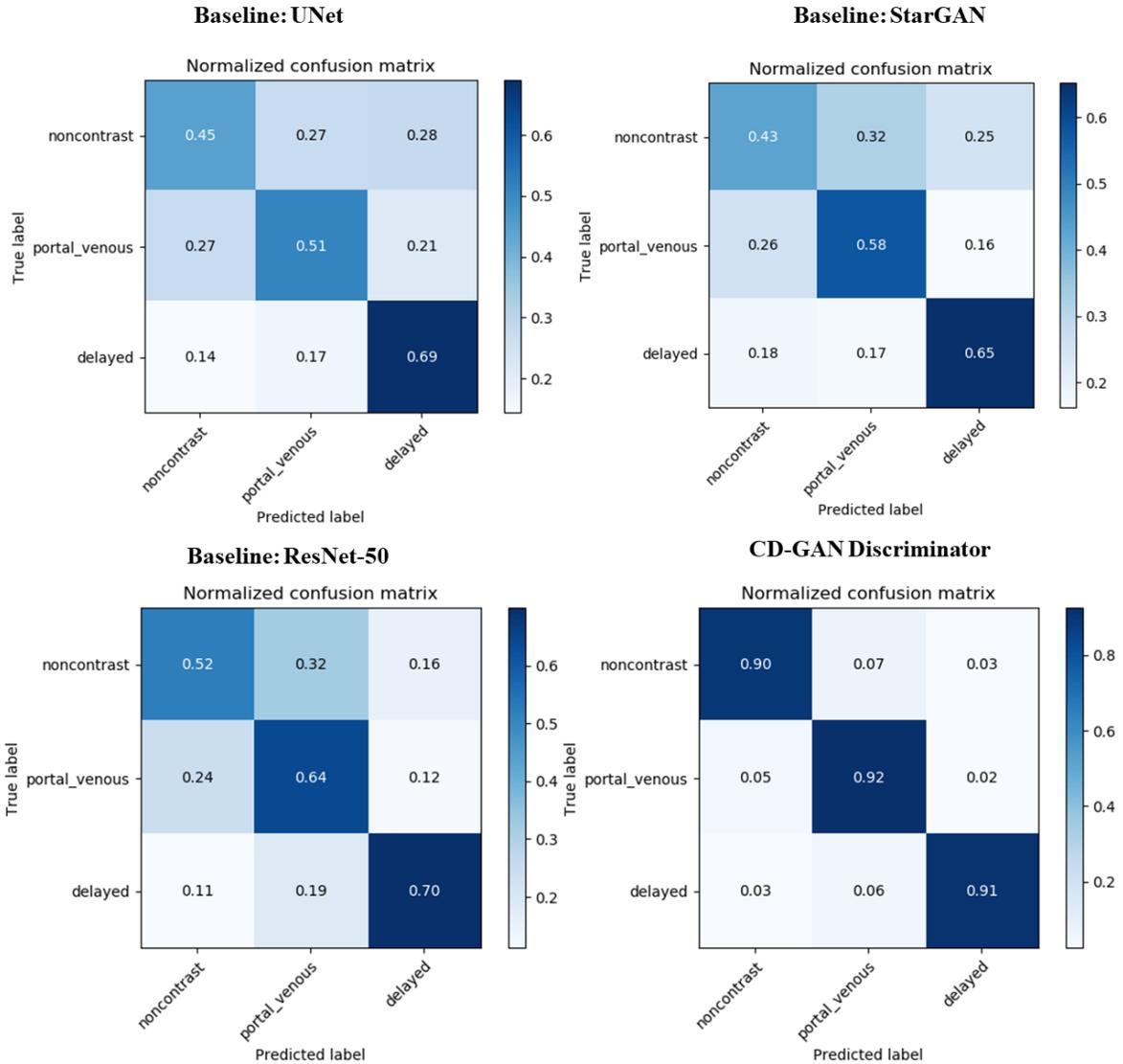

Figure 3. Slicecase confusion matrix result on withheld 30 paired subjects of baseline UNet (top left), ResNet50 (bottom left), StarGAN (top right) and proposed CD-GAN discriminator (bottom right). CD-GAN with adversarial loss and classification on real/synthetic images boosted the general performance with significant improvement ($p < 0.001$ with paired t test) overall accuracy score in CD-GAN and state-of-the-art ResNet50.

## 3.2 Qualitative evaluation

As seen in Figure 3, CD-GAN classified difficult slices into correct categories. Errors are reviewed in Figure 4. The top left slice is an image near the pelvis region. In most cases, contrast materials will not be diffused to this region, only small part of kidney could be seen to distinguish the enhancement. The top middle slice is from a patient with fluid surrounding the liver and spleen so that it is hard to visually recognize the changes of intensity distribution with contrast materials. The top right slice is from heart area. Note that the delayed phased is a late enhancement phase where typically heart will no longer bright. On the bottom row, the left example is mis-predicted as non-contrast, then correct to delay phase as the delayed phased is enhanced in urinary system which is hard to visually recognize. The middle and right slices are representative incorrect predictions as the portal venous phase and delayed phase are similarly distinct from the kidneys and urinary system.

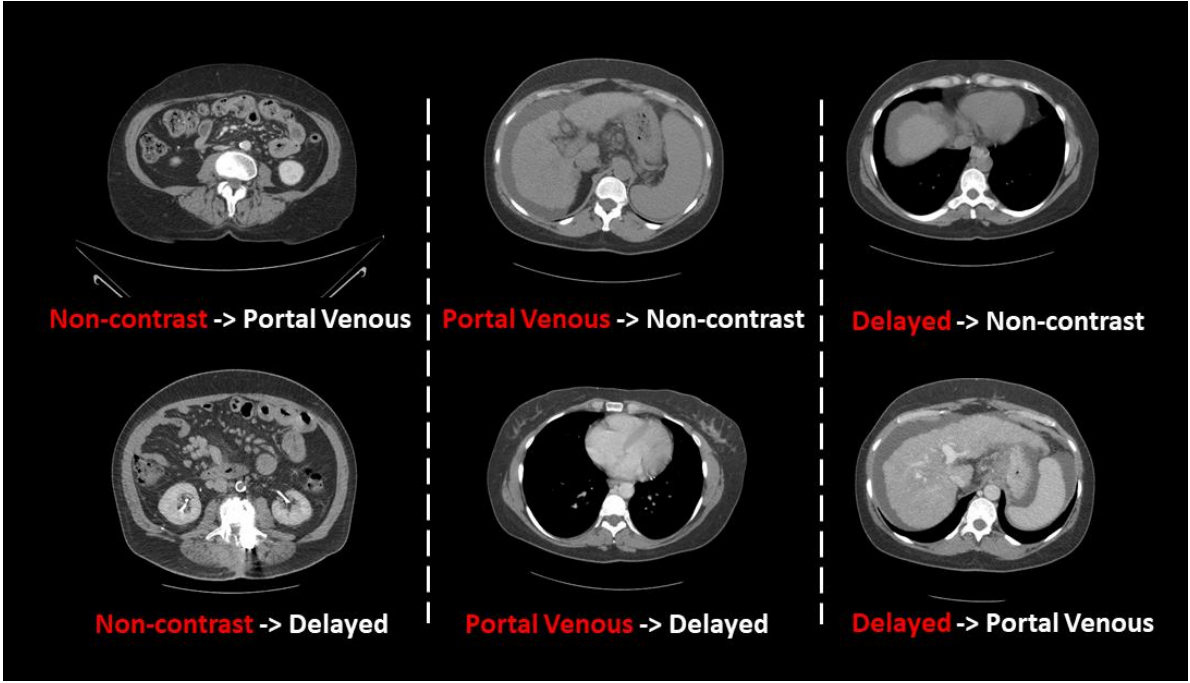

Figure 4. Representative qualitative erroes. Left: misclassification with non-contrast (red: ResNet50) corrected to right phase (white: CD-GAN Discriminator); Mid: mis-prediction of portal venous phase corrected to non-contrast and delayed phase; Right: error output of delayed phase corrections.

## 4. CONCLUSION AND DISCUSSION

Herein, we proposed a discriminator network for contrast disentangling GAN and provided robust classifier for distinguishing contrast CTs in clinical archives. The proposed method enables classification by using disentangling GAN with multi-domain discriminator inspired by StarGAN. Our CD-GAN discriminator has two novelties. First, the discriminator used adversarial loss and classification loss on both real/ synthetic images instead of canonical losses in single classification model. Second, the synthetic images are reconstructed by a generator with two steps, encoder disentangled the enhancement to an intermediate representation, and decoder added contrast phase identity on the representation to synthesis the synthetic image. We implemented each real slice with three synthetic targets, these images improved robustness and the overall performance of discriminator. The algorithm has been designed as an adversarial learning and representation learning framework. Additional work is needed to investigate possible applications of the synthetic result.

**Acknowledgement** This research is supported by Vanderbilt-12Sigma Research Grant, NSF CAREER 1452485, NIH grants, 2R01EB006136, 1R01EB017230 (Landman), and R01NS09529.This study was in part using the resources of the Advanced Computing Center for Research and Education (ACCRE) at Vanderbilt University, Nashville, TN. We gratefully acknowledge the support of NVIDIA Corporation with the donation of the Titan X Pascal GPU used for this research. The imaging dataset(s) used for the analysis described were obtained from ImageVU, a research resource supported by the VICTR CTSA award (ULTR000445 from NCATS/NIH) and Vanderbilt University Medical Center institutional funding. ImageVU pilot work was also funded by PCORI (contract CDRN-1306-04869).